\documentclass[aps,twocolumn,floatfix,superscriptaddress]{revtex4-1}
\usepackage[english]{babel}

\usepackage{amsmath,amsfonts,amssymb} 
\usepackage{physics}   

\usepackage{lmodern}
\usepackage{placeins}

\usepackage{graphicx}
\usepackage{color}
\usepackage{macros-math}

\usepackage{varioref}
\usepackage{cleveref}
\usepackage{soul}

\begin{document}

\title{Critical slowing down 
in driven-dissipative Bose-Hubbard lattices}

\author{Filippo Vicentini}
\affiliation{Universit\'{e} Paris Diderot, Sorbonne Paris Cit\'{e}, Laboratoire Mat\'{e}riaux et Ph\'{e}nom\`{e}nes Quantiques, CNRS-UMR7162, 75013 Paris, France}

\author{Fabrizio Minganti}
\affiliation{Universit\'{e} Paris Diderot, Sorbonne Paris Cit\'{e}, Laboratoire Mat\'{e}riaux et Ph\'{e}nom\`{e}nes Quantiques, CNRS-UMR7162, 75013 Paris, France}

\author{Riccardo Rota}
\affiliation{Universit\'{e} Paris Diderot, Sorbonne Paris Cit\'{e}, Laboratoire Mat\'{e}riaux et Ph\'{e}nom\`{e}nes Quantiques, CNRS-UMR7162, 75013 Paris, France}

\author{Giuliano Orso}
\affiliation{Universit\'{e} Paris Diderot, Sorbonne Paris Cit\'{e}, Laboratoire Mat\'{e}riaux et Ph\'{e}nom\`{e}nes Quantiques, CNRS-UMR7162, 75013 Paris, France}

\author{Cristiano Ciuti}
\email{cristiano.ciuti@univ-paris-diderot.fr}
\affiliation{Universit\'{e} Paris Diderot, Sorbonne Paris Cit\'{e}, Laboratoire Mat\'{e}riaux et Ph\'{e}nom\`{e}nes Quantiques, CNRS-UMR7162, 75013 Paris, France}
\date{\today}
 
 \begin{abstract}
We theoretically explore the dynamical properties of a first-order dissipative phase transition in coherently driven Bose-Hubbard systems, describing, e.g., lattices of coupled nonlinear optical cavities. Via stochastic trajectory calculations based on the truncated Wigner approximation, we investigate the dynamical behavior as a function of system size for 1D and 2D square lattices in the regime where mean-field theory predicts nonlinear bistability. We show that a critical slowing down emerges for increasing number of sites in 2D square lattices, while it is absent in 1D arrays. We characterize the peculiar properties of the collective phases in the critical region.
 \end{abstract}

\maketitle

\section{Introduction}
In a closed manybody quantum system at zero temperature, the pure ground state may undergo a quantum phase transition when there is a competition between two physical processes described by non-commuting Hamiltonian terms \cite{sachdev2001quantum}. In an open system  \cite{BreuerBook07}, the competition between unitary Hamiltonian evolution and dissipation can induce a dissipative phase transition for the steady-state in the thermodynamic limit \cite{KesslerPRA12}, as it has been discussed theoretically for photonic systems \cite{CarmichaelPRX15,MendozaPRA16,CasteelsPRA16,BartoloPRA16,CasteelsPRA17,CasteelsPRA17-2,Foss-FeigPRA17,BiondiarXiv16,BiellaarXiv17,SavonaarXiv17}, lossy polariton condensates \cite{Sieberer13,Sieberer14,Altman15} and spin models \cite{Lee13,Jin16,Maghrebi16,Rota17,OrusNat17}. 

Photonic systems are particularly promising to investigate dissipative phase transitions described by Bose-Hubbard-like models \cite{CarusottoRMP13,NohReview17,HartmannReview17,AMO2016934}, particularly in platforms based on patterned semiconductor microcavities \cite{JacqminPRL2014} and nonlinear superconducting microwave resonators \cite{HouckNP2012,Fitzpatrick17,LabouviePRL2016}. A few interesting experiments on photonic systems have been reported recently, such as a spectroscopic and dynamical study of a one-dimensional array where the first resonator is coherently pumped \cite{Fitzpatrick17} and the observation of a dissipative phase transition in a coherently-driven semiconductor micropillar \cite{RodriguezPRL2017,FinkarXiv17} or in a single superconducting nonlinear resonator \cite{FinkPRX2017}. 
The field is still in its infancy, but comprehensive experimental investigations of controlled one- and two-dimensional \cite{JacqminPRL2014} nonlinear photonic lattices are within reach.

In a lattice of coupled resonators with local boson-boson interaction $U$ a coherent and homogeneous driving of all the sites can create a macroscopic population of bosons in the zero-wavevector mode ($k = 0$). Being delocalized in space, the latter experiences a self-interaction of strength $U/N$, $N$ being the number of sites. If one retains only the $k=0$-mode operators in the driven-dissipative Bose-Hubbard model, such crude approximation predicts  a first-order phase transition for a critical driving strength \cite{CasteelsPRA17-2}. An interesting and challenging problem is to understand how the presence of the other modes with ${\bf k} \neq {\bf 0}$ affects the dynamics of the system. In particular, the emergence of criticality might depend on fluctuations associated to this multitude of modes and on the dimensionality of the lattice.  A recent work \cite{Foss-FeigPRA17} has reported calculations of the steady-state population for lattices  as a function of the driving strength, suggesting the presence of a first-order discontinuity in two-dimensional lattices, while only a smooth crossover in one-dimensional arrays. 

The finite-size scaling of the dynamical properties have not been systematically explored in driven-dissipative lattice systems. This is a challenging theoretical problem because it requires the study of a large number of modes and very long time scales. In equilibrium systems, a critical slowing down of transient dynamics is observed at the critical point when the Hamiltonian energy gap vanishes. Instead, in dissipative systems  a critical slowing down is expected to be related to the spectrum of the Liouvillian superoperator governing the time evolution of the density matrix \cite{KesslerPRA12}.  However, a clear demonstration of critical slowing down as a function of lattice size is still missing.
 
In this Letter, we explore the dynamical properties of the driven-dissipative Bose-Hubbard model both in 1D and 2D square lattices. Within the truncated Wigner approximation, we solve stochastic Langevin equations for the lattice fields and determine the dependence of relaxation time dynamics towards the steady-state as a function of lattice size. We are able to determine the presence of critical slowing down in 2D lattices due to the emergence of a first-order phase transition between a collective low-density and high-density phase. We characterize this paradigm of dissipative phase transition via a comprehensive study of the main observables.

\section{The driven-dissipative Bose-Hubbard Model}

\begin{figure}[t!]
	\includegraphics[width=0.8 \linewidth]{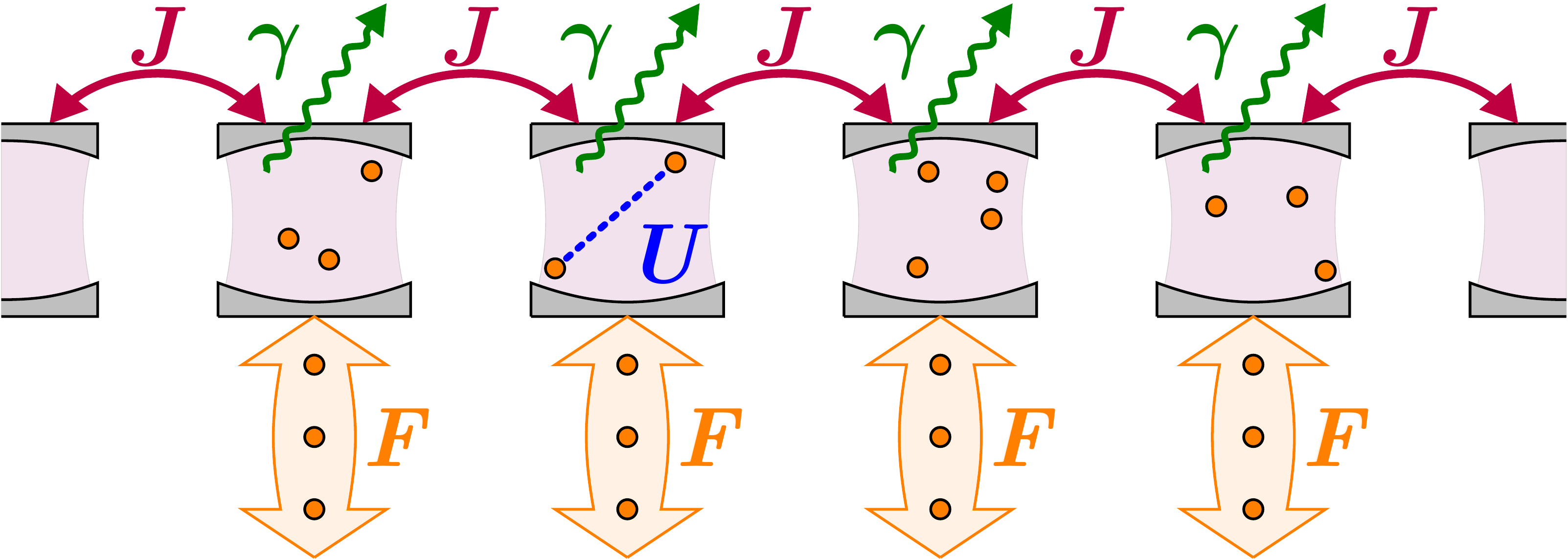}
	\vspace{0.1pt}
	\includegraphics[width=1 \linewidth]{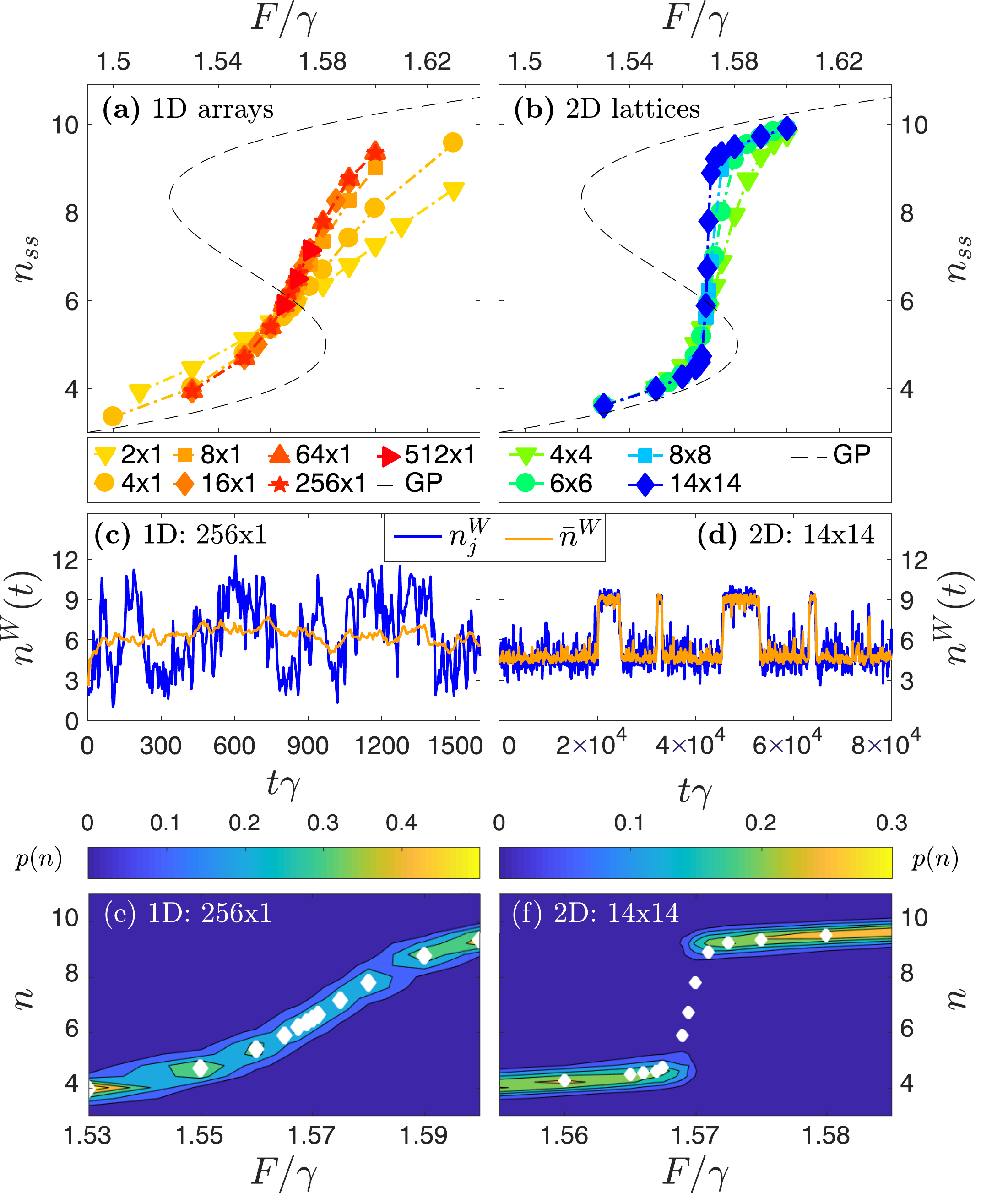}
	\caption{Top panel: the considered driven-dissipative Bose-Hubbard system is depicted (only the 1D case is shown). Left panels are for 1D arrays, while right panels refer to 2D square lattices. (a) and (b): steady-state average population per site versus driving amplitude $F$ (in units of the dissipation rate $\gamma$) for lattices of different size.  The dashed line is the prediction of the Gross-Pitaevskii mean-field theory.  
	(c), (d):  time-dependent single-trajectory population $n_j^W$in the j-th site (dark blue) and same quantity averaged over all sites $\bar{n}_j^W$ (light orange)  for $F=1.5695 \gamma$. (e), (f): contour plot of the probability distribution $p(n)$ of the site-averaged steady-state population versus the driving. White diamonds represent the steady-state average population per site, also shown in panels (a) and (b).
	(c) and (e) are for a $256 \times 1$ array, while (d) and (f) are for a $14 \times 14$ lattice.
	Trajectories have been computed via the truncated Wigner approximation with parameters: $U = 0.1 \gamma$, $\Delta = 0.1 \gamma$ and $zJ= 0.9 \gamma $ (hopping rate times the coordination number $z$). 
		\label{density}}
\end{figure}

The Bose-Hubbard model in presence of coherent driving with frequency $\omega_p$ is described by the following Hamiltonian (in the frame rotating with the drive, $\hbar = 1$):
\begin{equation}
	\oH = \sum_{j}-\Delta\oad_j\oa_j + \frac{U}{2}\hat{a}^{\dagger  2}_j\oa_j^2 + F\left(\oad_j+\oa_j\right) - J\sum_{<j,j'>} \oad_j\oa_{j'} 
	\label{eq:hamiltonian}
\end{equation}

where $\Delta=\omega_p-\omega_c$ is the detuning between the driving frequency and mode frequency $\omega_c$, $U$ the on-site interaction, $F$  the homogeneous driving field (the phase is chosen in such a way that $F$ is real) and $J$ the hopping coupling between two nearest neighboring sites (see \cref{density} top panel). In the following, $z$ will denote the number of nearest neighbors ($z = 2$ and $z =4$ respectively for the 1D and 2D lattices considered in this work).

To describe the dissipative dynamics, we will consider the following Lindblad master equation for the lattice reduced density matrix $\orho$, assuming an uniform Markovian single-boson loss rate $\gamma$ \cite{BreuerBook07}:
\begin{equation}
	\derivate{\orho}{t} =  \lind\orho = -i\comm{\oH}{\orho} + \frac{\gamma}{2}\sum_j\left[2\oa_j\orho\oad_j  - \acomm{\oad_j\oa_j}{\orho} \right].
	\label{eq:lindblad-me}
\end{equation}

 The Liouvillian non-hermitian superoperator $\lind$ has a complex spectrum of eigenvalues $\{ {\lambda_r} \}$ with ${\rm Re}(\lambda_r)\leq 0$, defined by the eigenvalue equation $\lind \hat{\rho}_{r} =  \lambda_{r} \hat{\rho}_{r}$. The steady-state is usually unique \cite{AlbertPRA14} and corresponds to the zero eigenvalue. The real part of the non-zero eigenvalues is responsible for the transient relaxation of the density matrix to the non-equilibrium steady-state $\hat{\rho}_{ss}$. The slowest relaxation dynamics is due to the eigenvalue with the smallest real part (in absolute value). We call  $\lambda = \text{min}_r \,\vert \text{Re}(\lambda_r) \vert $ the Liouvillian frequency gap, which is the inverse of the asymptotic decay rate towards the steady-state. A dissipative phase transition is expected to be characterized by a critical slowing down associated to the closing of the Liouvillian gap in the thermodynamic limit \cite{KesslerPRA12}.
 
In this work,  we explore lattices in a regime where the so-called truncated Wigner approximation method can be applied \cite{VogelPRA89,CarusottoPRB2005,CarusottoRMP13}.  In general, the Lindblad master equation can be mapped exactly into a third-order differential equation for the quasi-probability Wigner function, which is a representation of the density matrix. In the limit of small $U$, the third-order derivatives can be neglected so that the differential equation \eqref{eq:lindblad-me} becomes a Fokker-Planck equation \cite{carmichael1998statistical} for a well defined probability function \cite{VogelPRA89,CarusottoRMP13}. 
The latter can be solved via a stochastic Montecarlo approach \cite{Rackauckas17} described by a set of Langevin equations for the complex field $\alpha_j(t)$ of the boson mode in the $j$-th site:
\begin{equation}
\begin{split}
	\dot{\alpha_j} = \left[ -i (\Delta -  U (\vert \alpha_j \vert ^2 - 1) - \gamma/2 )\right] \alpha_j - \\ -i J \sum_{j'}  \alpha_{j'}  + i F + \sqrt{\gamma/2} \,\chi (t)
	\label{eq:TruncatedWignerSDE},
\end{split}
\end{equation}
where $j'$ runs over the nearest neighbors of $j$ and $\chi(t)$ is a normalized random complex gaussian noise such that $\expval{\chi(t)\chi(t')}=0$ and $\expval{\chi(t)\chi^*(t')}=\delta(t-t')$.
Within this formalism, expectation values for symmetrized products of operators \cite{VogelPRA89,CarusottoRMP13} are obtained by averaging over different stochastic trajectories through the relation $\expval{\acomm{(\hat{a}_i^\dagger)^n}{\hat{a}_j^m}_s} = \frac{1}{N_\text{traj}}\sum_{r}({\alpha}_{i,r}^*)^n \alpha_{j,r}^m $, where the index $r$ runs over the $N_\text{traj}$ random trajectories. 

\begin{figure}[t!]\includegraphics[width=\linewidth]{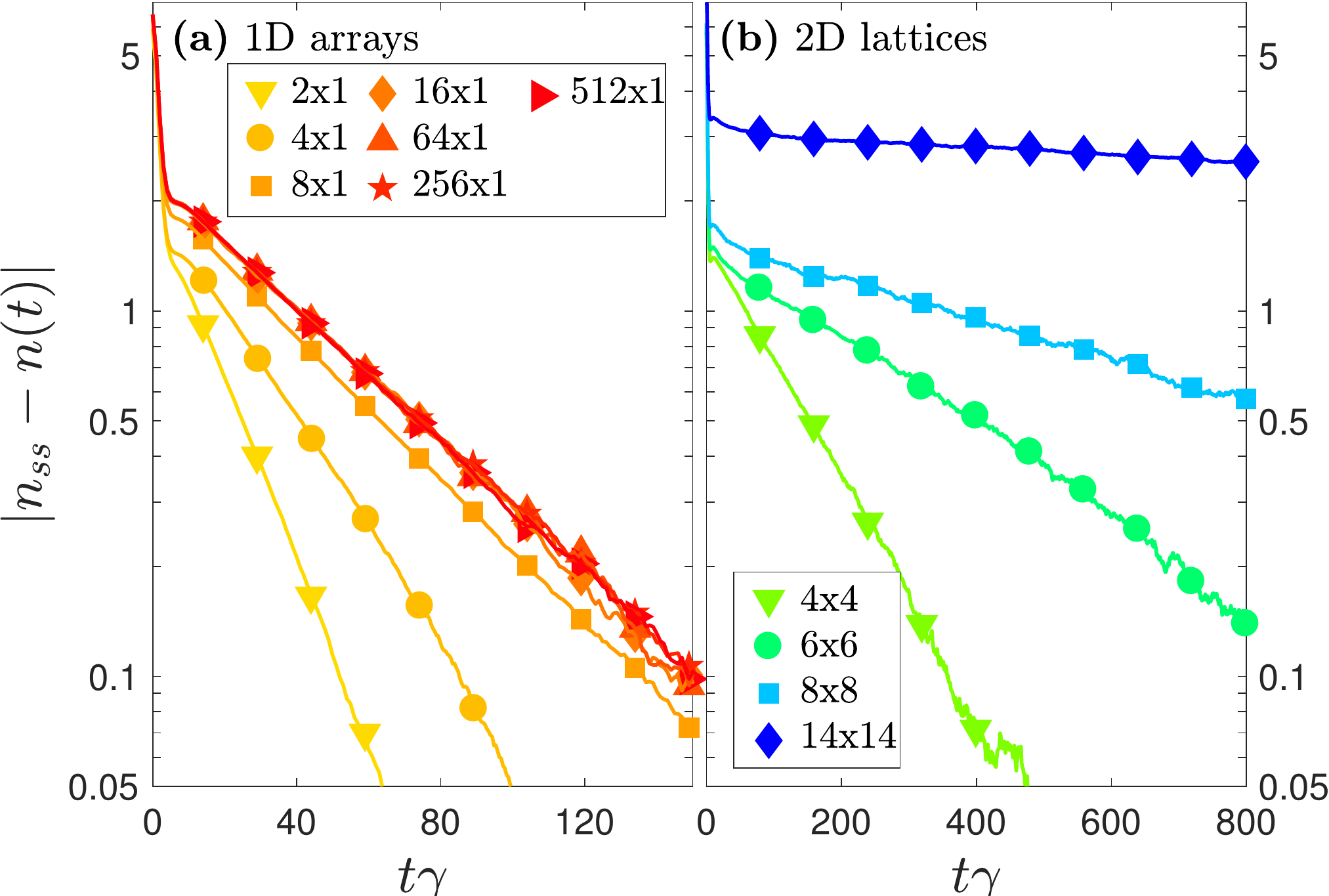}
	\caption{Transient dynamics of the absolute difference between the mean occupation number $n(t)$ and its steady-state value $n_{ss}$ for 1D arrays (a) and 2D lattices (b) of different sizes, with driving amplitude $F=1.57 \gamma$. Other parameters as in \cref{density}.}
	\label{fig:dynamics_n}
\end{figure}

\section{Critical behaviour in the bistable region}

Here we will explore the driven-dissipative Bose-Hubbard model and investigate a first-order phase transition in a regime where mean-field theory predicts bistability. Within a Gross-Pitaevskii-like mean-field approach \cite{LeBoitePRA14}, the master equation for the lattice density matrix is replaced by a simple equation for the mean-field $\alpha_j = \langle \hat{a}_j \rangle$, which is the same as \cref{eq:TruncatedWignerSDE}, but without the noise terms. In the homogenous case ($\alpha_j = \alpha$), the steady-state equation takes the nonlinear form $\vert \alpha \vert^2 ((\Delta + zJ - U \vert \alpha \vert^2)^2 + \gamma^2/4) = F^2$, which can have three non-degenerate solutions for a given $F$,  two of which are dynamically stable. As in all mean-field theories \cite{LeBoitePRL13,LeBoitePRA14,BiondiarXiv16}, the effect of hopping depends only on $z J$, with the lattice dimension playing no role. Hence, in the following, when comparing 1D versus 2D lattices, we will consider the same value of $z J$, so that differences will only be due to effects beyond mean-field.

In \cref{density}(a) we present results obtained with the truncated Wigner approximation for the steady-state site-averaged population  $n_{ss} = \frac{1}{N} \sum_{i=1}^N  {\rm{Tr}} (\hat{\rho}_{ss} \hat{a}_i^{\dagger} \hat{a}_i)$ for 1D arrays of different length $L$ (up to $L = 512$).  
In \cref{density}(b), the same observable is reported for 2D $L \times L$ lattices (up to $14 \times 14$). Both 1D and 2D calculations have been performed with periodic boundary conditions.
For the value $U = 0.1 \gamma$ considered in the following, we have successfully benchmarked (see \cref{app:Benchmark}) the accuracy of the truncated Wigner approximation for small lattices by comparison with brute-force numerical integrations of the master equation and also calculations based on the corner-space renormalization method \cite{FinazziPRL15}. In both \cref{density}(a) and (b) the Gross-Pitaevskii-like mean-field prediction is depicted by the dashed line. 
While, in general, mean-field theories exhibit multistability, the density-matrix solution of the master equation is under quite general assumptions unique \cite{DrummondJPA80,AlbertPRA14}: indeed, quantum fluctuations make the mean-field solutions metastable so that on a single trajectory the system switches back and forth from one metastable state to another on a time scale related to the inverse Liouvillian gap \cite{VogelPRA89,CasteelsPRA16,WilsonPRA16,RodriguezPRL2017} (see also \cref{density}(c)). 
 The results in \cref{density}(a) show that the S-shaped multivalued curve of the mean-field theory is replaced by a single-valued function, which depends on the array size $L$.
Remarkably, by increasing the size $L$ of the array $n_{ss}(F)$ eventually converges to a curve with a finite slope. On the other hand, in 2D the slope of $n_{ss}(F)$ does not saturate when increasing the size $L$ of the lattices, suggesting the emergence of a discontinuous jump in the thermodynamic limit compatible with a first-order phase transition. 
\begin{figure}[t!]\includegraphics[width=0.98\linewidth]{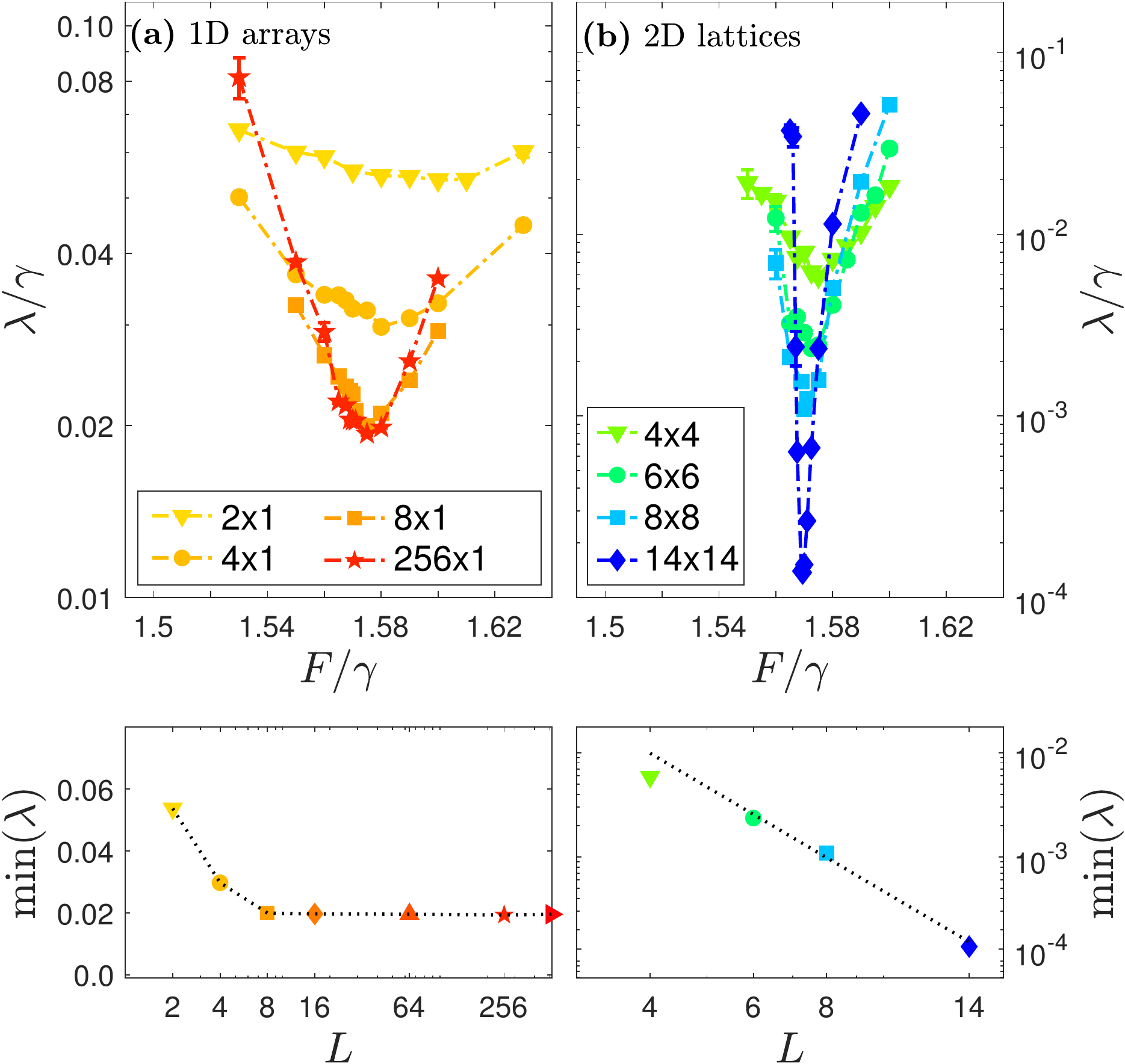}
	\caption{The Liouvillian gap $\lambda$ (log scale) versus the driving amplitude for several $L  \times 1$ arrays (a) and $L \times L$ lattices (b). Notice the different scales used for the 1D and 2D case. The insets show the minimum of $\lambda$ as a function of the size $L$. Error bars are within the symbol size.
		Parameters as in \cref{density}.} 
	\label{gap}		
\end{figure}

In \cref{density}(c) and \cref{density}(d), we present the dynamics of the boson population in a single stochastic Wigner trajectory for the 1D and 2D lattices, respectively. In the considered regime of interaction $U$, Wigner trajectories have a direct correspondence to local oscillator measurements \cite{drummond2014quantum},  such as those carried out via homodyne detection techniques \cite{wiseman2010quantum, gardiner2004quantum}. In 1D, switches between the two metastable mean-field solutions are barely visible in the population of the $j$-th site $n_{j}^W(t)$ (blue curve) and absent in the site-averaged population $\bar{n}^W(t)=1/N \sum_{j=1}^N n^W_j(t)$ (orange curve), consistent with the formation of moving domains with low and high density inside the array \cite{Foss-FeigPRA17}.
On the contrary, the 2D lattice exhibits a strikingly different behavior, with a clear random switching behavior of $n_{i}^W(t)$ between two well definite metastable states. The populations in all sites switch collectively since $n_{j}^W(t)$ and $\bar{n}^W(t)$ strongly overlap.
Furthermore, notice that the 2D timescales are far longer than in the 1D case, indicating a significantly slower dynamics.
A particularly insightful  quantity is the probability number distribution $p(n)$ defined as follows. We consider a time $t_s$ where the system has reached the steady state and statistically collect all the values of $n = \bar{n}^{W}  (t > t_s)$ for all the considered trajectories.
 The results for $p(n)$ are presented in \cref{density}(e,f) for different values of the driving amplitude $F$. We notice that, in the 1D case, this distribution is monomodal for all values of $F$ and the steady-state mean value of the population follows the peak of this distribution. In the 2D lattice $p(n)$ exhibits a completely different behavior: it has a single peak in the limit of small and large $F$, while it is bimodal in proximity of the critical region. Here, for finite-size the steady-state expectation value $n_{ss}$ falls in a region of negligible probability ($p(n_{ss}) \simeq 0$) in-between two peaks corresponding to the low and high population phases. When the 2D lattice size is increased, the crossover between the two phases becomes steeper and therefore the bistable region also becomes narrower, eventually collapsing to a single point when $L\rightarrow\infty$. This explains why in large lattices a fine scan in $F$ is necessary to observe this feature.

To investigate the emergence of criticality in the dynamical properties,  we calculated the time evolution towards the steady-state value $n_{ss}$ of the site-averaged mean occupation number $n(t)= \frac{1}{N} \sum_{i=1}^N  {\rm{Tr}} (\hat{\rho}(t) \hat{a}_i^{\dagger} \hat{a}_i)$ , taking the vacuum as initial state. For values of $F$ close to the critical point,  $n(t) -n_{ss}$ decays exponentially to zero at large times as reported in \cref{fig:dynamics_n}. In this asymptotic regime, the dynamics is dominated by the Liouvillian gap $\lambda$, which can be extracted by fitting the results with $n(t) = n_{ss} + A e^{- \lambda t}$. Note that in order to have enough accuracy, calculations have required up to $10^6$ stochastic Wigner trajectories for each data point. Experimentally, the asymptotic decay rate can be also measured using the time-dependence of the second-order correlation function \cite{FinkarXiv17}, dynamical optical hysteresis \cite{RodriguezPRL2017} and switching statistics \cite{Fitzpatrick17,RodriguezPRL2017}.
The particular case of $F=1.57 \gamma$ is analyzed in \cref{fig:dynamics_n}, where we plot $|n_{ss} - n(t)|$ for 1D arrays (panel a) and 2D lattices (panel b) of different sizes. 
For this fixed value of $F$, the dynamics gets slower as the size of the simulated system is increased. While in the 1D case the exponential decay rate saturates in the thermodynamic limit, this is not the case for 2D systems. The emergence of critical slowing down is quantified in \cref{gap}, where we provide the size-dependence of the Liouvillian gap $\lambda$ versus $F$.
In \cref{gap}(a), we report results for 1D arrays: it  is apparent that, when the size $L$ is large enough, the Liouvillian gap converges to a finite value for all the values of $F$, thus proving the absence of critical slowing down.
The behavior is strikingly different for 2D lattices, as shown in \cref{gap}(b): in this case, every curve $\lambda(F)$ presents a minimum, which becomes smaller and smaller
when the size $L$ of the lattice is increased.  As shown in the inset of  \cref{gap}(b), the minimum of $\lambda$ follows the power-law decay  $\min_{\lambda}(L)\propto L^{-\eta}$, with exponent $\eta = 3.3 \pm 0.1$.  Since the phase transition is of first order, this exponent is not universal \cite{CasteelsPRA17-2,FinkarXiv17}. To verify this, we computed the critical exponent in lattices with a different nonlinearity (the other parameters were unchanged), finding $\eta=5.3\pm 0.1$ for $U/\gamma=0.06$ and $\eta=1.7\pm0.2$ for $U/\gamma=0.15$ (see the Appendix).

\begin{figure}[t]
	\includegraphics[width=1 \linewidth]{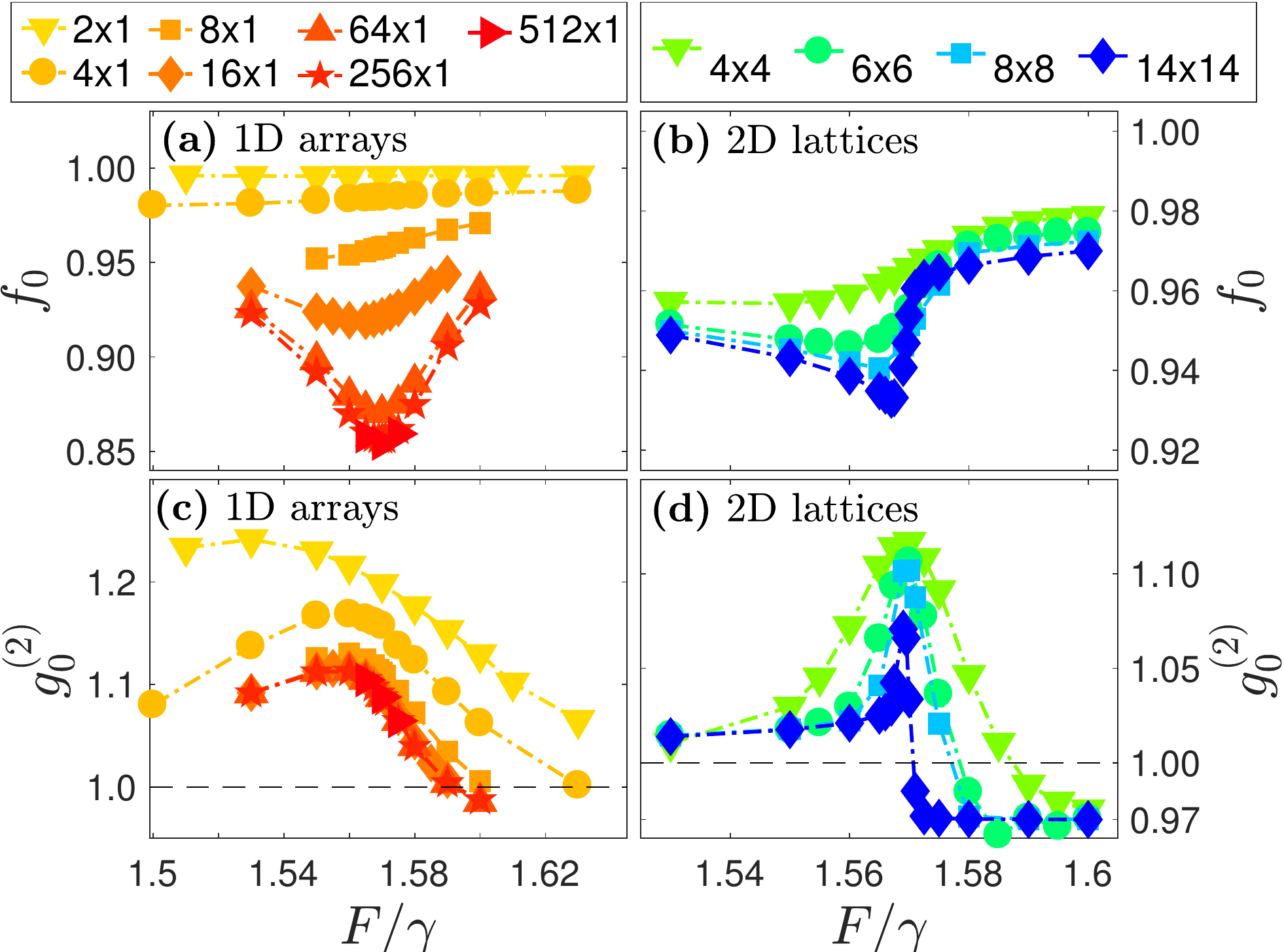}
	\caption{(a) and (b): population fraction $f_0 = n_{k=0}/{n_{tot}}$ in the zero-momentum mode as a function of the driving amplitude. 
	(c) and (d): zero-delay local second-order correlation $g^{(2)}_0$ versus $F$. 
	Left panels are for 1D arrays, right panels for 2D lattices. Same parameters as in \cref{density}. } 
	\label{observables}		
\end{figure}

The phase transition observed here in 2D lattices is reminiscent of what predicted analytically in the driven-dissipative Bose-Hubbard model through an approximation where only the $k=0$-mode is retained \cite{CasteelsPRA17-2}. Therefore, one may expect that a macroscopic population in the $k=0$ mode would always give rise to a critical behavior.  In this regard, we studied the fraction $f_0 = n_{k=0}/{n_{tot}}$ of bosons in the $k=0$-mode, where $n_{k=0}$ is the steady-state population of the driven $k=0$-mode and $n_{tot}$ is the total lattice population. In  \cref{observables}(a) and (b) we report the finite-size analysis of $f_0$ as a function of $F$. In the region of mean-field bistability, $f_0$ presents a minimum in both 1D and 2D. In 1D this minimum saturates to a finite value as one approaches the thermodynamic limit, while in 2D $f_0$  exhibits a behavior consistent with a finite jump at the critical point. For the considered interaction, in both cases the population of the driven mode is dominant  ($f_0$  close to $1$), showing that the fluctuations induced by the coupling to non-zero momentum modes destroy the critical behavior in 1D. 

Lastly, we present the local equal-time second-order correlation function $g^{(2)}_0=  \langle \oad_j\oad_j\oa_j\oa_j  \rangle / \langle \oad_j \oa_j  \rangle^2$ as a function of $F$. This quantity describes the amplitude of the fluctuations in the field, and has been employed extensively to investigate critical behavior in in optical systems. In 1D this quantity has a broad peak whose shape is shown to converge for large enough $L$ (\cref{observables}(c)), while in 2D (\cref{observables}(d)) the finite-size results show an emerging  singular behavior in its derivative at the critical point. The same qualitative behavior is also observed in the large population limit of a single-mode nonlinear resonator \cite{CasteelsPRA17-2,BartoloPRA16}, which is equivalent to the $k=0$ approximation described above.

\section{Conclusions}

In conclusion, we have theoretically predicted the critical slowing down associated to a dissipative transition in the driven-dissipative Bose-Hubbard model.
We have revealed the emergence of critical dynamics in 2D lattices via a finite-size analysis,  which is instead absent in 1D arrays, indicating that the lower critical dimension for this non-equilibrium model is $d < 2$. 
We have shown that in 1D arrays fluctuations destroy criticality of the dynamics even if the driven mode is macroscopically occupied.
The asymptotic decay rate associated to the Liouvillian frequency gap has been measured in nonlinear photonic systems with different techniques \cite{RodriguezPRL2017,Fitzpatrick17,FinkarXiv17}, hence the critical slowing down predicted here as a function of lattice size is within experimental reach and can unveil fundamental properties of dissipative phase transitions. Many intriguing studies can be foreseen at the horizon, including the role of disorder as well as the critical behavior of exotic open photonic lattices with geometric frustration \cite{MukherjeePRL15,BabouxPRL16,CasteelsPRA16Lieb,BiondiPRL15} or quasi-periodicity \cite{TanesePRL14,AMO2016934}. 

\acknowledgements{We would like to thank N. Bartolo, A. Biella, J. Bloch, W. Casteels, N. Carlon Zambon, M. Foss-Feig for discussions. We acknowledge support from ERC (via Consolidator Grant CORPHO No. 616233).}

\appendix
\section{Benchmark of the Truncated Wigner Approximation}
\label{app:Benchmark}

\begin{figure}[t]
		\begin{center}
			\vspace{0.3cm}
			\includegraphics[width=0.9\linewidth]{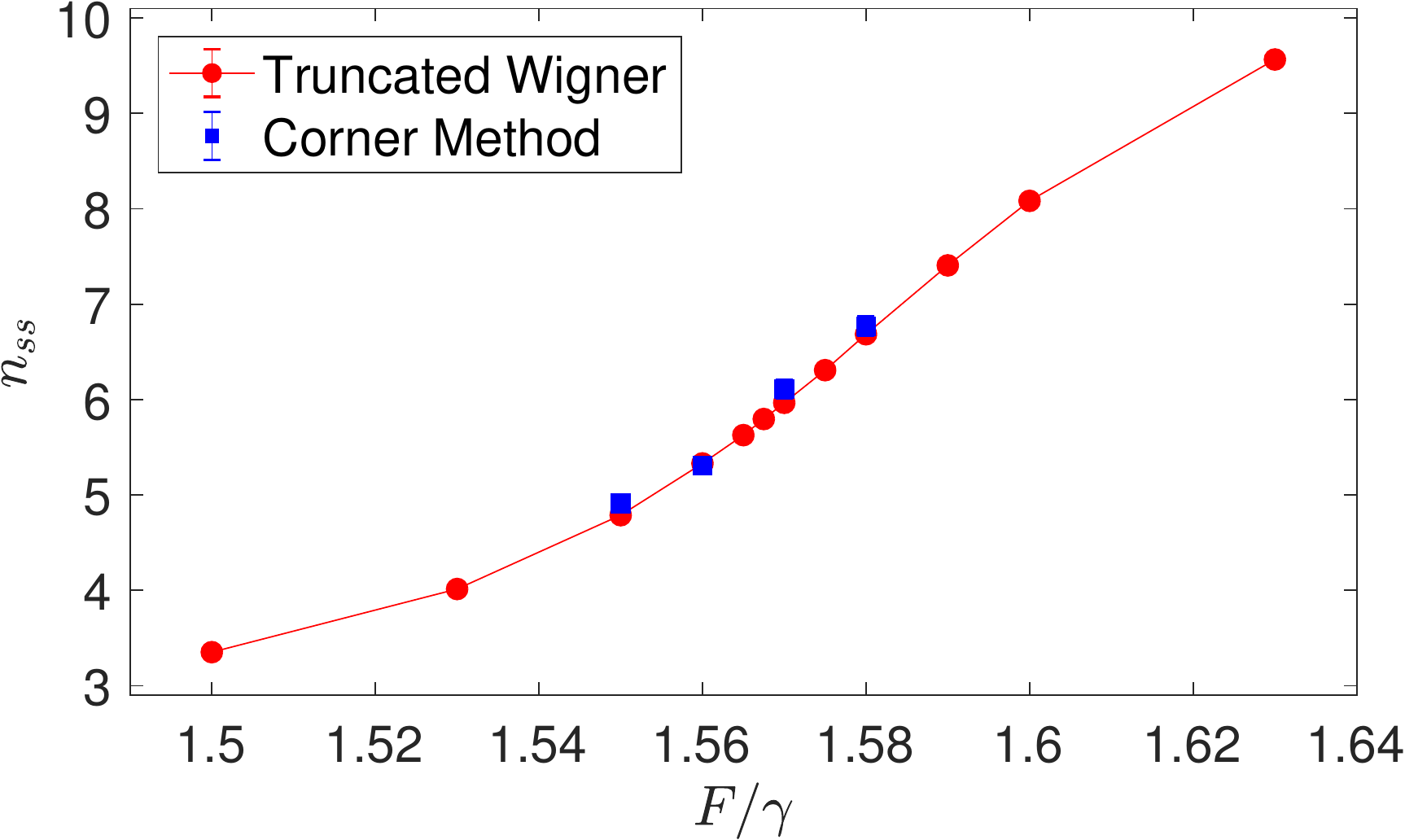}
			\caption{Steady-state average boson occupation per site as a function of the driving $F/\gamma$ in a $4\times 1$ array: different symbols correspond to different numerical methods.  The statistical error is of the order of the symbol size.  Parameters are $U/\gamma=0.1$, $zJ/\gamma=0.9$, $\Delta/\gamma=0.1$. }
			\label{fig:comparison_nss}
		\end{center}
\end{figure}
\begin{figure}[!t]
	\begin{center}
	\vspace{0.3cm}
		\includegraphics[width=0.9\linewidth]{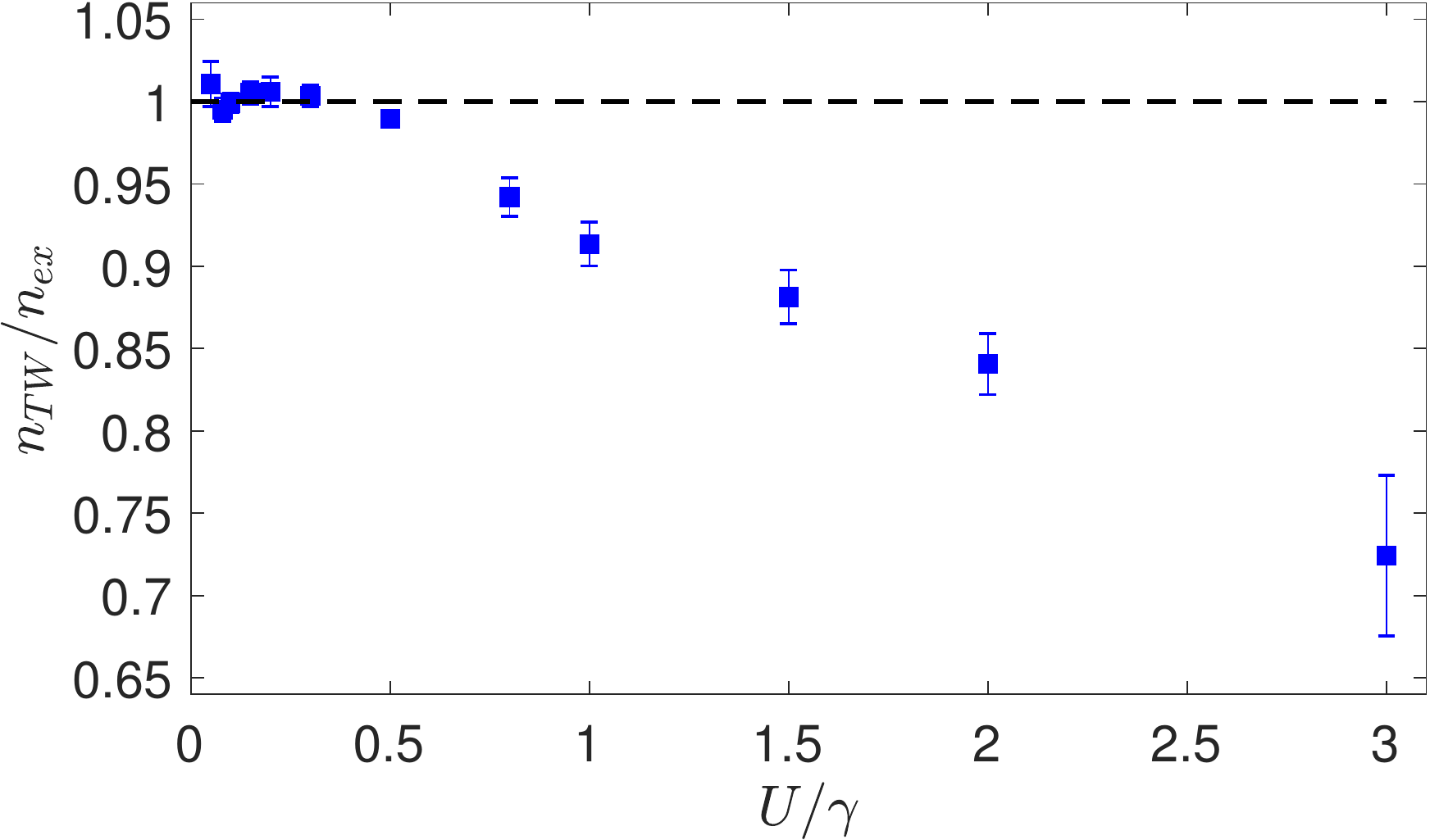}
		\caption{Ratio between the steady-state average occupation obtained through the Truncated Wigner approximation $n_{TW}$ and the exact Runge-Kutta Integration of the Lindblad Master Equation $n_{ex}$ in a $2\times 1$ array. The error bars refer to the statistical noise of the results associated to the stochastic Langevin simulations. $F/\gamma$ has been varied so that $UF^2/\gamma^3=2.465$ is kept constant; $zJ/\gamma=0.9$ and $\Delta/\gamma=0.1$ are fixed. Note that the results for $U/\gamma \leq 0.2$ have been obtained with the corner-space renormalization.}
		\label{fig:comparison_U}
	\end{center}
\end{figure}

\begin{figure}
	\begin{center}
		\includegraphics[width=0.9\linewidth]{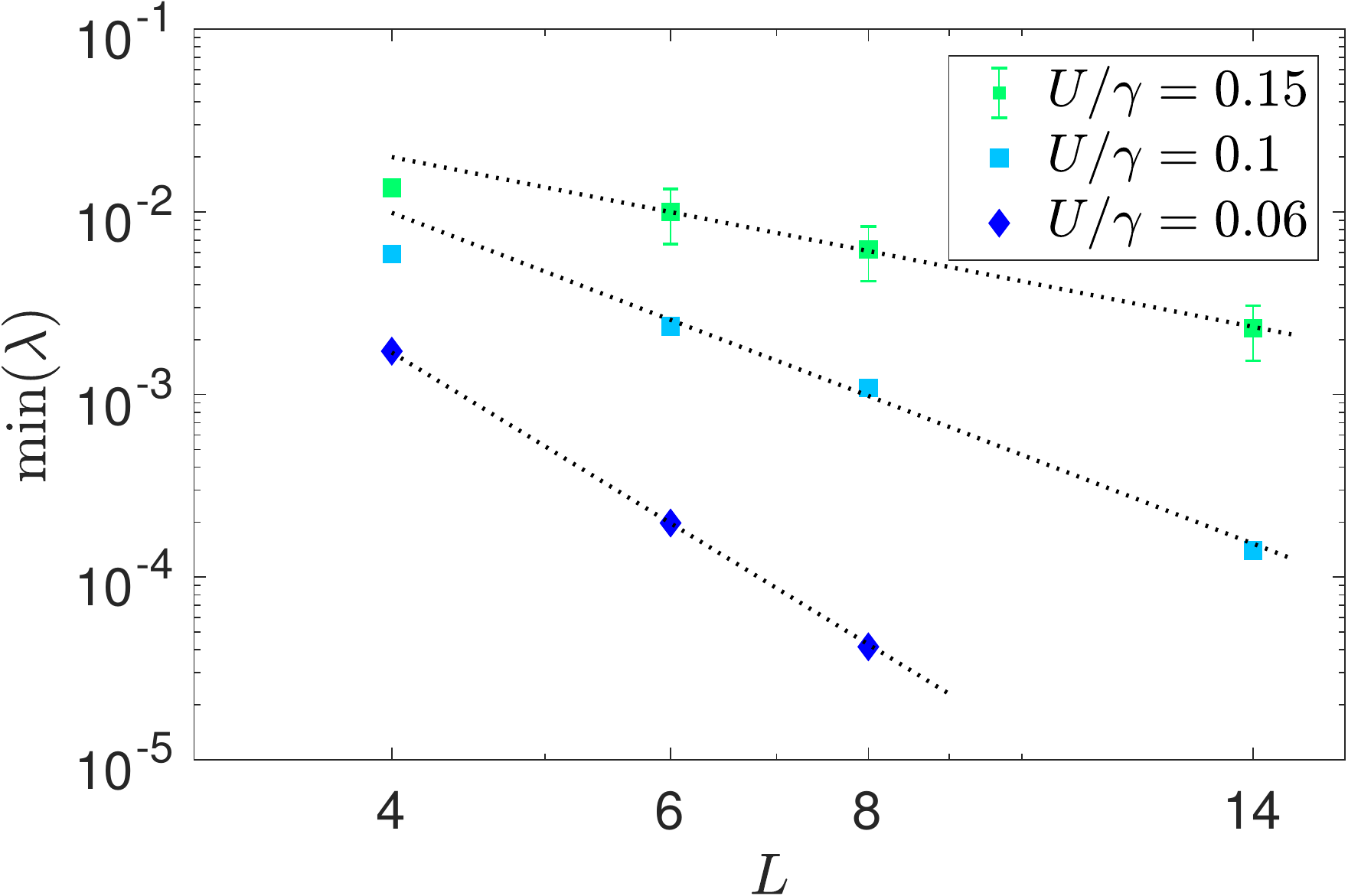}
		\caption{Minimum of the Liouvillian gap $\lambda$ as a function of the size $L$ of 2D lattices, for different values of  $U/\gamma$. The critical exponent is $\eta=1.7\pm0.2$ for $U/\gamma=0.15$, $\eta=3.3\pm0.1$ for $U/\gamma=0.1$ and $\eta=5.3\pm 0.1$ for $U/\gamma=0.06$. Parameters:  $zJ/\gamma=0.9$ and $\Delta/\gamma=0.1$.}
		\label{fig:comparison_gap}
	\end{center}
\end{figure}

In this appendix, we present numerical results showing that the Truncated Wigner Approximation is accurate in the regime of parameters considered in the manuscript.
To do so, we compare its results to what was obtained with numerically exact methods for small systems. Moreover, we show how the power-law decay of the Liouvillian gap changes when the normalized interaction $U/\gamma$ is varied.

In \cref{fig:comparison_nss}, we present the steady-state average population in a $4\times 1$ array computed with the Truncated Wigner approximation and with the corner-space renormalization method \cite{FinazziPRL15} finding an excellent agreement between the two.

The values considered are the same as in the main text. We would like to point out that for the considered value of $U/\gamma=0.1$, a brute-force integration of the master equation for a one-site system requires a cutoff of $N_{max}=40$ bosons in order to achieve adequate numerical convergence. In a $4\times 1$ lattice the required dimension of the Hilbert space would be $40^4=2.56\cdot 10^6$ which cannot be handled numerically without more advanced methods. For the parameters considered in the main text, this lattice can still be tackled by the corner-space renormalization method (going to larger lattice sizes would require significantly larger computational resources).

In \cref{fig:comparison_U} we present the ratio $n_{TW}/n_{ex}$ between the steady-state average population obtained via the Truncated Wigner approximation $n_{TW}$ and exact methods $n_{ex}$ as a function of the nonlinearity $U/\gamma$. We used this quantity to identify the range of values in $U/\gamma$ for which the Truncated Wigner approximation is quantitatively accurate, finding that for $U/\gamma\leq0.3$ the Truncated Wigner yields results within $1\%$ of the exact value.

In \cref{fig:comparison_gap} we present the minimum of the Liouvillian gap $\lambda$ as a function of lattice size $L$ for several 2D lattices with different nonlinearities. We find that the power-law exponent increases as $U/\gamma$ is decreased.

\bibliographystyle{apsrev4-1}
\bibliography{bibliography}

\end{document}